\pdfoutput=1

\documentclass{aastex631}

\usepackage{amssymb,amsmath}
\usepackage{soul}
\usepackage{bm}

\newcommand{\changes}[1]{#1}

\begin{document}

\title{A Classic Nonlinearity Correction Algorithm for Detectors Read Out Up-The-Ramp}

\correspondingauthor{Timothy D.~Brandt}
\email{tbrandt@stsci.edu}
\author[0000-0003-2630-8073]{Timothy D.~Brandt}
\affiliation{Space Telescope Science Institute \\ 3700 San Martin Drive \\ Baltimore, MD 21218, USA}



\begin{abstract}

We derive an algorithm for computing a classic nonlinearity correction---applicable to constant and uniform illumination---in the presence of read noise and photon noise.  The algorithm operates simultaneously on many nondestructive ramps at a range of count rates and directly computes the function transforming measured counts into linearized counts.  We also compute $\chi^2$ for the corrected ramps, enabling the user to identify the polynomial degree beyond which $\chi^2$ ceases to improve significantly.  The computational cost of our algorithm is linear in the number of reads and ramps, reaching $\sim$100 hours to derive a correction for all $4096 \times 4096$ pixels of a Hawaii-4 RG detector from 186 illuminated 55-read ramps on a 2023 Macbook Pro laptop ($\approx$10,000 reads per pixel).  We identify a potential source of bias in the nonlinearity correction when combining ramps of very different illuminations, together with effective mitigations.  We apply our algorithm to a random set of pixels from the Roman Space Telescope's Wide Field Instrument.  We find that a $\geq$9$^{\rm th}$ order nonlinearity correction is needed, at which point $\chi^2$ is close to its theoretically expected value and beyond which $\chi^2$ improves little with increasing polynomial order.  Python software implementing our algorithm is available at \url{https://github.com/RomanSpaceTelescope/SOCReferenceFileCode}. 

\end{abstract}



\section{Introduction} \label{sec:introduction}

An ideal detector illuminated at a constant intensity will accumulate charge linearly with time.  Apart from photon noise and read noise, it will record a digital signal that increases with time at a constant rate.  A real detector will exhibit a number of nonideal effects.  Some of these couple neighboring pixels together, while others are specific to individual pixels.  Examples of the former include interpixel capacitance \citep[IPC,][]{Moore+Ninkov+Forrest_2004} and the brighter-fatter effect \citep[BFE,][]{Guyonnet+Astier+Antilogus+etal_2015,Plazas+Shapiro+Smith+etal_2017}.  Examples of the latter include a nonlinear relationship between charge and voltage on a pixel's capacitor and nonlinearities in the analog-to-digital converter that converts measured voltage into a digital number, or DN.  

We will use the term ``classic nonlinearity'' to refer to the combined impact of several nonideal detector effects that make a pixel's accumulated counts a nonlinear function of time \citep{McCaughrean_1988}.  Classic nonlinearity, as we define it in this paper, applies to a uniformly illuminated detector \citep{Vacca+Cushing+Rayner_2004}.  It does not include effects like IPC that couple neighboring pixels together.  It also does not include count rate dependent nonlinearity, in which the slope of a pixel's accumulated counts is a nonlinear function of the illumination level even when the accumulated counts themselves are a linear function of time.  A classic nonlinearity correction is simply intended to make the accumulated counts a linear function of time under constant, uniform illumination.

A classic nonlinearity correction will give the true counts $z_i$ in read $i$ as a function of the observed counts $y_i$ as
\begin{equation}
    z_i = f(y_i)
    \label{eq:general_nonlinearity}
\end{equation}
where $f$ is some function.  A polynomial approximation to $f$ is often used for convenience \citep{Vacca+Cushing+Rayner_2004,Robberto_2011,2017jwst.rept.5167C}, though other functional forms could also be used.  Equation \eqref{eq:general_nonlinearity} implicitly assumes that the corrected signal is a function of the measured signal only, with no dependence on the signal history.  As such, classic nonlinearity will not correct for effects like persistence or burn-in that apply when the illumination changes.  In persistence, charge continues to accumulate after illumination decreases or ceases, while in burn-in, charge initially accumulates more slowly as charge traps in the pixel capture some of the initial photoelectrons and the pixel acclimates to a higher illumination \citep{Smith+Zavodny+Rahmer+Bonati_2008}.

A nonlinearity correction may be derived differently in a CCD, which can be read out only once before being reset, and a detector that can be read out nondestructively.  For a CCD we must either make many measurements at different intensities with known intensity ratios, or we must make many measurements at the same intensity but with varying exposure times \citep{Gilliland+Brown+Kjeldsen+etal_1993,Baldry_1999}.  For a detector that can be read out many times before being reset, nonlinearity can be seen, and measured, within a single set of reads (termed a ramp).  A nonlinearity correction may thus be derived from a single exposure \citep{Robberto_2011,2017jwst.rept.5167C}. 

This paper considers only the classic nonlinearity correction, and we treat the case of a detector read out many times nondestructively.  We assume that there is a parametrized function $f$ (we use a polynomial) satisfying Equation \eqref{eq:general_nonlinearity}.  We then build a statistical framework to derive the coefficients describing $f$ and to measure its goodness-of-fit.  Because we limit ourselves to classic nonlinearity, we use only data taken at temporally constant and spatially uniform illumination.  

We structure the paper as follows.  Section \ref{sec:generalapproach} outlines the problem and our approach.  In Section \ref{sec:formalism} we derive our nonlinearity correction algorithm assuming a polynomial correction function and a covariance matrix representing read noise and photon noise.  Section \ref{sec:numerical} discusses the numerical and computational performance of our approach.  Section \ref{sec:syntheticdata} evaluates its performance on synthetic data, while Section \ref{sec:comparison_canipe} compares its performance to that of the legacy \cite{2017jwst.rept.5167C} approach.  In Section \ref{sec:romanresults} we apply our algorithm to thermal vacuum test data from the Roman Space Telescope's Wide Field Instrument detectors \citep{Mosby+Rauscher+Bennett+etal_2020}.  We conclude with Section \ref{sec:conclusions}.

\section{Statement of the Problem and General Approach} \label{sec:generalapproach}

\begin{figure}
    \centering\includegraphics[width=0.5\textwidth]{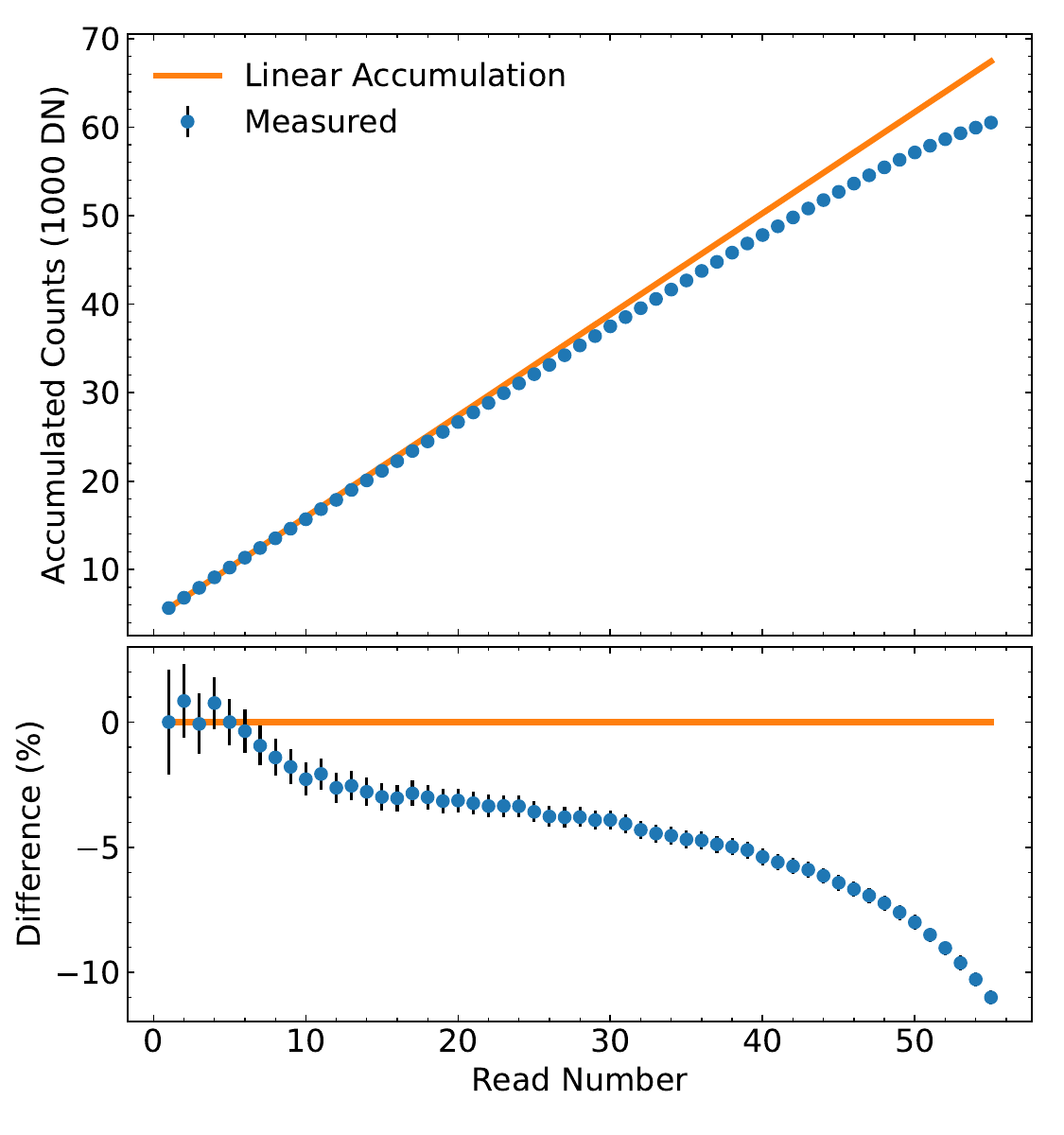}
    \caption{Example of the nonlinear response of a Roman-WFI pixel to constant illumination.  The orange line shows a constant count rate, while the blue points show the actual measurements.  The lower panel shows the difference between the blue points and the orange line normalized by the orange line's accumulated charge from reset. The errors represent photon noise and are highly correlated. \label{fig:example_nonlinearity}}
\end{figure}

As noted in Section \ref{sec:introduction}, a real detector will record a nonlinear signal with time even when subject to constant illumination.  Figure \ref{fig:example_nonlinearity} shows the response of a pixel on the Roman Space Telescope's Wide Field Instrument \citep[Roman-WFI,][]{Domber+Gygax+Aumiller+etal_2022, Schlieder+Barclay+Barnes+etal_2024} detector array \citep{Mosby+Rauscher+Bennett+etal_2020} to constant illumination.  The accumulated counts as a function of time do not follow a straight line, but fall below the count level expected based on an extrapolation from the first few reads.  

Under constant, uniform illumination, a nonlinearity correction transforms the measured counts $y$ into corrected counts $z$ according to Equation \eqref{eq:general_nonlinearity}.  In the context of Figure \ref{fig:example_nonlinearity}, the goal of a classic nonlinearity correction is to bring the blue points in Figure \ref{fig:example_nonlinearity} into statistical agreement with a straight line.  In this section we adopt a polynomial approximation for $f$, so that 
\begin{equation}
    z_i = \sum_{j=0}^N a_j y_i^j .
\end{equation}
The degree of the polynomial $N$ and the values of the coefficients $\{a_j\}$ specify the nonlinearity correction.  Astronomically important measurements consist of count rates, so the $a_0$ term does not matter.  If we adopt the convention that the nonlinearity correction is unity for a small number of counts, then we may also take $a_1=1$.  The task at hand is then to use many reads across one or more exposures to derive the coefficients $\{a_j\}$ given a desired polynomial order $N$.  Ideally, such an analysis would account for the covariance matrix of the accumulated counts $y_i$ at the various reads $i$.  This covariance matrix is far from diagonal: a photon recorded in a given read is also recorded in all subsequent reads.

A common way of approaching the problem of deriving a nonlinearity correction is to solve a similar, but different, problem.  A ramp measures counts as a function of time, and a polynomial may be fit to this assuming a diagonal covariance matrix for the accumulated counts at the various reads.  That polynomial may then be approximately inverted into another polynomial.  This second polynomial constitutes the nonlinearity correction that renders the ramp into something very close to a straight line.  Another approach is to first fit for the ``ideal'' count rate $\eta$ by fitting a polynomial to the counts as a function of time $t$ and evaluating the slope $\eta$ at $t=0$.  Assuming that 
\begin{equation}
    z_i = \eta t + z_0,
\end{equation}
we now have theoretical values for the corrected counts $\{z_i\}$.  A nonlinearity correction may then be fit to derive the polynomial coefficients $\{a_j\}$. 

Neither of the approaches outlined above adopts the correct covariance matrix for the reads or solves for the linearity correction directly.  The latter approach, in particular, requires deriving a ``true'' count rate from the first few reads and assuming it to hold throughout the exposure.  This is especially risky when combining exposures at potentially different illumination levels.  For this latter reason NIRCam nonlinearity corrections are derived from long ramps that are averaged read-by-read before being analyzed to derive a nonlinearity correction \citep{2017jwst.rept.5167C}.  

In this paper we will reframe the problem slightly to use a very close approximation to the correct covariance matrix and to directly compute the forward, rather than the inverse, linearity correction.  We will assume that the true, or corrected, accumulated counts are a linear function of time,
\begin{equation}
    z_i = b t_i ,
\end{equation}
where $b$ is the (unknown) incident rate of detectable photons at the average reset value $y_0$, and $t_i$ is the (known) time since reset of read $i$.  We then have
\begin{equation}
    b t_i = y_0 + y_i + \sum_{j=2}^N a_j y_i^j .
    \label{eq:nonlinearity_eq1}
\end{equation}
This may be thought of as a Taylor expansion centered on $y=0$.  We will then compute the differences $y_{i+1} - y_i$ so that the covariance matrix is tridiagonal \citep{Brandt_2024a}.  We will also assume that the covariance matrix between $z_i$ and $z_{i-1}$ is the covariance matrix described in \cite{Brandt_2024a} based on read noise and photon noise (i.e.~that photon noise properly applies to the corrected counts rather than to the raw counts).  This assumption only matters at very high counts where read noise is irrelevant.  

Taking the $t_i$ to be consecutive integers (so that the count rate $b$, itself a nuisance parameter, has units of counts per read), we can recast Equation \eqref{eq:nonlinearity_eq1} as
\begin{equation}
    b = (y_{i + 1} - y_i) + \sum_{j=2}^N a_j (y_{i + 1}^j - y_i^j) .
    \label{eq:nonlinearity_eq2}
\end{equation}
We will now recenter the Taylor expansion around a reference value $y_0$ (which could, in practice, be a superbias or dark level) and also allow the slope of the linearity correction at $y=y_0$ to be free by adding a coefficient $a_1$ to the first term.  This will give me more flexibility in imposing constraints on $b$; we can always rescale the polynomial correction to $a_1=1$ later.  We then have
\begin{equation}
    b t_i = {\rm constant} + \sum_{k=1}^N a_k \left(y_i  - y_0\right)^k
\end{equation}
or, in the space of read differences,
\begin{equation}
    b = \sum_{k=1}^N a_k ((y_{i + 1} - y_0)^k - (y_i - y_0)^k) \label{eq:b_eq1}
\end{equation}
where $N$ is the degree of the polynomial correction and $i = 1, \ldots n$ is the read.  Because the expansion is recentered, the coefficients $a_k$ in Equation \eqref{eq:b_eq1} differ from the $a_k$ in Equation \eqref{eq:nonlinearity_eq2}.  The rest of the paper proceeds from Equation \eqref{eq:b_eq1}.  

\section{Formalism} \label{sec:formalism}

Section \ref{sec:generalapproach} recasts the problem of fitting a nonlinearity correction in an equivalent way using read differences.  In this section, we use the covariance matrix of the read differences to create $\chi^2$ from Equation \eqref{eq:b_eq1}.  We then combine the $\chi^2$ values of many ramps to derive a single nonlinearity correction from all of them.  The values $\{ a_k \}$, i.e., the nonlinearity correction itself, will be the same for all of these ramps, but the value of $b$ will vary from one ramp to the next.  In other words, the various ramps used will have different levels of illumination.  

To compute the nonlinearity correction coefficients $\{ a_k \}$, we will define and then minimize $\chi^2$ summed over all ramps.  For notational convenience, we will define the quantity $g_{i,k}$ as
\begin{equation}
    g_{i,k} = \left(y_{i + 1} - y_0\right)^k - \left(y_i - y_0\right)^k.
    \label{eq:g_ik}
\end{equation}
Equation \eqref{eq:g_ik} represents the difference between successive reads raised to the $k$-th power after subtracting a reference value; $g_{i,1} = d_i$ in \cite{Brandt_2024a}.  

For a single ramp, we will write $\chi^2$ as
\begin{equation}
    \chi^2 = 
    \begin{bmatrix}
    \sum a_k g_{n-1,k} - b \\
    \sum a_k g_{n-2,k} - b \\
    \vdots \\
    \sum a_k g_{1,k} - b \\
    \end{bmatrix}^T {\bf C}^{-1}
    \begin{bmatrix}
    \sum a_k g_{n-1,k} - b \\
    \sum a_k g_{n-2,k} - b \\
    \vdots \\
    \sum a_k g_{1,k} - b \\
    \end{bmatrix} 
    \label{eq:chisq_oneramp}
\end{equation}
where ${\bf C}^{-1}$ is the inverse of the covariance matrix of the read differences comprising the ramp, and is given in \cite{Brandt_2024a}.  The $\chi^2$ that we would like to optimize is given by Equation \eqref{eq:chisq_oneramp} summed over all ramps.  

\changes{We next} rewrite the single-ramp $\chi^2$ of Equation \eqref{eq:chisq_oneramp} as a quadratic form in the coefficients $\{a\}$ and $b$.  The quadratic form only has nonzero coefficients for the quadratic terms (e.g.~$b^2$, $a_2 a_1$, etc.); \changes{there are no constant or linear terms. We have} 
\begin{equation}
    \chi^2 = \sum_{k} \sum_l a_k a_l {\cal A}_{kl}
\end{equation}
\changes{with}
\begin{equation}
    {\cal A}_{kl} =
    \begin{bmatrix}
    g_{n-1,k} \\
    g_{n-2,k} \\
    \vdots \\
    g_{1,k}
    \end{bmatrix}^T {\bf C}^{-1}
    \begin{bmatrix}
    g_{n-1,l} \\
    g_{n-2,l} \\
    \vdots \\
    g_{1,l}
    \end{bmatrix} ,
    \label{eq:chisq_oneterm}
\end{equation}
taking $g_{i,0}=-1$ and letting $a_0$ correspond to $b$. \changes{Thus, ${\cal A}_{i,0}$ is the coefficient of $a_i b$, and ${\cal A}_{0,0}$ is the coefficient of $b^2$.}

\changes{Equation \eqref{eq:chisq_oneterm}} is similar to Equation (59) of \cite{Brandt_2024a}, where the differences of reads are replaced by the differences of reads raised to a power (note that \cite{Brandt_2024a} mistakenly omits $\beta$ from their Equation (61)).  These terms may be efficiently computed using the same approach as in \cite{Brandt_2024a}.  Writing this out as an explicit sum, we have
\begin{align}
    {\cal A}_{kl} = 
    \sum_{i=1}^{n-1} g_{i,k} \sum_{j=1}^{n-1} g_{j,l} \left( {\rm C}^{-1} \right)_{ij} .
\end{align}
where $i$ and $j$ index read number, $n$ is the number of reads, and $n-1$ is the number of read differences.  The second, interior sum is nearly identical to Equation (4) of \cite{Brandt_2024b}.  In close correspondence to that paper, we replace $d_i$ with $g_{i,k}$ in Equations (51) and (74) of \cite{Brandt_2024a} to define $(\Theta G)_{i,k}$ and $(\Phi G)_{i,k}$ in place of $(\Theta D)_{i}$ and $(\Phi D)_{i}$.  We then have, from \cite{Brandt_2024b},
\begin{equation}
    {\cal A}_{kl} = \sum_{i=1}^{n-1} g_{i,k} \frac{(-1)^i}{\theta_n} \left(\phi_{i+1} \left( \Theta G \right)_{i,l} + \theta_{i-1} \left(\Phi G\right)_{i,l}\right) .
    \label{eq:chisq_element}
\end{equation}
The computational cost of computing each ${\cal A}$ term is linear in the number of reads.  

With the computations outlined above, we may rewrite $\chi^2$ for this ramp as
\begin{equation}
    \chi^2 = \begin{bmatrix}
        a_{N} \\
        a_{N - 1} \\
        \vdots \\
        a_1 \\
        b \\
    \end{bmatrix}^T
        \begin{bmatrix}
        {\cal A}_{N,N} & {\cal A}_{N,N-1} & \ldots & {\cal A}_{N, 1} & {\cal A}_{N,0} \\ 
        {\cal A}_{N-1,N} & {\cal A}_{N-1,N-1} & \ldots & {\cal A}_{N-1, 1} & {\cal A}_{N-1,0} \\ 
        \vdots & \vdots & \ddots & \vdots & \vdots \\
        {\cal A}_{1,N} & {\cal A}_{1,N-1} & \ldots & {\cal A}_{1, 1} & {\cal A}_{1,0} \\ 
        {\cal A}_{0,N} & {\cal A}_{0,N-1} & \ldots & {\cal A}_{0, 1} & {\cal A}_{0,0} \\ 
    \end{bmatrix}
    \begin{bmatrix}
        a_{N} \\
        a_{N - 1} \\
        \vdots \\
        a_1 \\
        b \\
    \end{bmatrix} .
    \label{eq:chisq_expanded}
\end{equation}
This is now a $(N + 1) \times (N + 1)$ matrix, where the computational cost of computing each term is ${\cal O}(n)$.  If we have $m$ ramps, then the matrix becomes $(N + m) \times (N + m)$, with Equation \eqref{eq:chisq_expanded} embedded within this larger matrix and the results summed over all ramps.  The cross terms, \changes{the coefficients of $b_v b_w$}, are zero between different ramps $v$ and $w$.  

Thus far, we have not imposed a constraint on the $b$ parameters.  They will differ from one ramp to the next assuming that the illumination varies due both to the need to achieve dynamic range and to imperfect stability in the setup.  If we freely fit all $b$ coefficients together with the nonlinearity coefficients $\{a\}$, the best-fit values of all parameters will be zero to produce $\chi^2 = 0$.  To avoid this, we will require that the sum of the count rates for all ramps, i.e.~the sum of all $b$ values, is fixed.  This constraint does not require a particularly exact value of $\sum b$ to be effective.  If $\sum b$ is slightly smaller than the sum of the true count rates, the linear term of the nonlinearity correction $a_1$ can absorb that difference by being slightly less than unity.  The nonlinearity correction may be rescaled later if desired.

In order to impose our constraint, we replace the $b$ constant from the final ramp (indexed by $m$) with a value of
\begin{equation}
    b_m = B_{\rm tot} - \sum_{j=1}^{m-1} b_j
    \label{eq:b_constraint}
\end{equation}
where $B_{\rm tot}$ is the (fixed) sum of all of the corrected ramp slopes and $b_j$ is the (unknown) count rate of ramp $j$.  In practice, we set $B_{\rm tot}$ to be the sum of the medians of the observed count rates over the first few reads for all ramps.  As noted above, if this value is too low, it will be compensated by a value of $a_1$ somewhat less than unity.  \changes{While we will impose the constraint of Equation \eqref{eq:b_constraint} by substitution, the same result could be obtained using a Lagrange multiplier.}

Our constraint removes one free parameter from the problem: the $\chi^2$ matrix becomes of dimension $(N + m - 1) \times (N + m - 1)$. The contribution to the $\chi^2$ matrix from the last ramp (indexed by $m$) is more complicated than the contributions from the other ramps, which are shown in Equation \eqref{eq:chisq_expanded}.  It may be derived from 
\begin{equation}
    \chi_m^2 = 
    \begin{bmatrix}
    \sum a_k g_{n-1,k} + \sum b_j - B_{\rm tot} \\
    \sum a_k g_{n-2,k} + \sum b_j - B_{\rm tot} \\
    \vdots \\
    \sum a_k g_{1,k} + \sum b_j - B_{\rm tot} \\
    \end{bmatrix}^T {\bf C}^{-1}
    \begin{bmatrix}
    \sum a_k g_{n-1,k} + \sum b_j - B_{\rm tot} \\
    \sum a_k g_{n-2,k} + \sum b_j - B_{\rm tot} \\
    \vdots \\
    \sum a_k g_{1,k} + \sum b_j - B_{\rm tot} \\
    \end{bmatrix} .
    \label{eq:chisq_lastramp}
\end{equation}
The ${\cal A}_{kl}$ terms are the same as for the other ramps, but the cross-term matrix entries \changes{corresponding to the coefficients of $b_v b_w$} are not zero.  Instead, they are all equal to what \changes{the coefficient of} $b^2$ would be if we were treating this ramp the same as all of the others, i.e.~to ${\cal A}_{0,0}$ from Equation \eqref{eq:chisq_expanded}: this part of the matrix is constant.  With nonzero cross-terms on the $\{b\}$ from the last ramp, every term in the $(N + m - 1) \times (N + m - 1)$ $\chi^2$ matrix is now nonzero.  

There is one final complication: Equation \eqref{eq:chisq_lastramp}, when expanded,
now contains terms proportional to the $\{a\}$ and $\{b\}$ coefficients to the first power.  These are linear terms that are also proportional to $B_{\rm tot}$, along with a constant term proportional to $B_{\rm tot}^2$.  As a result, the form of Equation \eqref{eq:chisq_expanded} no longer includes all terms.  The coefficients of the linear terms will be, e.g., $-2B_{\rm tot} {\cal A}_{k,0}$ for $a_k$ and $-2B_{\rm tot} {\cal A}_{0,0}$ for any of the $b$ terms, where the extra factor of two comes from the binomial expansion.  The computation of these coefficients does not require significant additional work.  We could ignore the constant term $B_{\rm tot}^2$ as it vanishes when differentiating $\chi^2$, but it is necessary if we want to compute $\chi^2$ itself to evaluate the overall goodness-of-fit.  It is given by $B_{\rm tot}^2 {\cal A}_{00}$.

Denoting the $(N + m - 1) \times (N + m - 1)$ matrix of quadratic coefficients defined by a larger version of the matrix in Equation \eqref{eq:chisq_expanded} by {\bf A} and the vector of linear coefficients by $-2 \bm{\xi}$ (so that $\bm{\xi}$ lacks the factor of $-2$ shown above), we have
\begin{equation}
    \chi^2 = 
    \begin{bmatrix}
        a_N \\ 
        a_{N-1} \\
        \vdots \\
        a_1 \\
        b_{m-1} \\
        \vdots \\
        b_1 \\
    \end{bmatrix}^T
    \left(
    {\bf A} \begin{bmatrix}
        a_N \\ 
        a_{N-1} \\
        \vdots \\
        a_1 \\
        b_{m-1} \\
        \vdots \\
        b_1 \\
    \end{bmatrix} 
    - 2 \bm{\xi}
    \right)
     + 
     B_{\rm tot} {\xi}_b . \label{eq:chisq_full}
\end{equation}
For the last term, we have replaced $B_{\rm tot}^2 {\cal A}_{00}$, with ${\cal A}_{00}$ evaluated from the last ramp, with $B_{\rm tot} {\xi}_b$, where ${\xi}_b$ refers to any of the (identical) entries of $\bm{\xi}$ that multiplies one of the $b$ terms.  Our next step is to set all of the partial derivatives of Equation \eqref{eq:chisq_full} equal to zero.  This gives
\begin{equation}
    {\bf A} \begin{bmatrix}
        a_N \\ 
        a_{N-1} \\
        \vdots \\
        a_1 \\
        b_{m - 1} \\
        \vdots \\
        b_1 \\
    \end{bmatrix}
     = 
     \bm{\xi} .
     \label{eq:linear_equation}
\end{equation}
This is now a linear system that may be straightforwardly solved to give the best-fit nonlinearity correction coefficients $\{a \}$.  If the formal covariance matrix of these coefficients is desired, it is available via the inverse of ${\bf A}$.  Should the user wish to impose that $a_1 = 1$, they may scale all of the $\{a\}$ coefficients  by the best-fit value of $a_1$ to compensate.  This is equivalent to scaling the $b$ nuisance terms and $B_{\rm tot}$ so that the linearized count rate equals the true count rate at $y=y_0$.  If the user wishes to evaluate $\chi^2$, for example to see how much it changes with increasing order of the correction or to evaluate the formal goodness-of-fit, they may evaluate Equation \eqref{eq:chisq_full} with the vector of parameters obtained by solving Equation \eqref{eq:linear_equation}.

The approach outlined in this section requires an estimate of the count rate in order to compute the covariance matrix for all of the reads.  We adopt a two-step iterative approach to this.  We first derive a crude estimate of the count rate from the median of count differences and fit for a nonlinearity correction together with all of the individual count rates.  We then use these count rates from the first iteration to recompute the covariance matrices and derive our final nonlinearity correction.  

Finally, the approach outlined in this section assumes that the read noise is the same for all of the reads after converting to the corrected counts, and that photon noise also properly applies to the corrected counts.  \changes{We assess the impact of this assumption in Section \ref{subsec:varyingcountrates}; we find it to be small.}

\section{Numerical Considerations, Implementation, and Computational Performance} \label{sec:numerical}

Assuming a large number of reads per ramp, much of the computational effort required for the nonlinearity correction is for the computation of the individual matrix elements of Equation \eqref{eq:chisq_expanded} using Equation \eqref{eq:chisq_element}.  The resulting matrix {\bf A} can be ill-conditioned if a high-order nonlinearity correction polynomial is to be fitted.  We adopt two approaches to overcome this problem:
\begin{enumerate}
    \item We transform the domain of measured counts, typically 0 to $2^{16}-1$ for unsigned 16-bit integers, to an interval close to $(-1, 1)$; and
    \item We (optionally) perform the fitting in the Legendre polynomial basis set rather than the standard polynomial basis.  This requires a redefinition of Equation \eqref{eq:g_ik} to refer to the Legendre polynomial of degree $k$.
\end{enumerate}

We baseline numerical performance using a set of 285 ramps, each 55 reads long, about half of which extend to digital saturation, restricted to a small region of one detector on Roman-WFI.  We use a gain of 1.8~e$^-$/DN and a single-read read noise of 5~DN to construct a covariance matrix for each ramp; these are typical values for the Roman-WFI detectors.  We then evaluate the condition number of the matrix ${\bf A}$ for a single pixel chosen at random.  Matrix inversion becomes problematic when the condition number approaches the inverse of floating point epsilon, or $\sim$10$^{15}$ for double precision.  For a tenth-order polynomial fit, we obtain a condition number for the matrix ${\bf A}$ of $\sim$10$^{11}$ for a polynomial in standard form and $\sim$10$^6$ using the Legendre basis.  In both cases, we are many orders of magnitude away from a point where numerical difficulties arise.  For a twentieth order polynomial, the condition numbers become $\sim$10$^{18}$ for standard form while remaining below 10$^7$ for Legendre basis.  As a result, we recommend the Legendre basis when fitting a nonlinearity polynomial correction of degree $\gtrsim$10.  Still, our nonlinearity corrections derived using the standard polynomial basis are relatively well-behaved even for the 20-th order correction.  Evaluating high-order polynomials is generally problematic, and artifacts do start to appear in our fits at very high orders ($\gtrsim$20) regardless of the polynomial basis used.  This is due to our subtraction of polynomials in Equation \eqref{eq:g_ik}, the extensive floating point operations in the evaluation of Equation \eqref{eq:chisq_element},  and the truncation error that inevitably follows.  

We implement our algorithm in Python, building heavily on the {\tt fitramp} package described in \cite{Brandt_2024a}.  We use the calculation of the covariance matrix for the reads directly from that work.  Rather than compute the best-fit coefficients of the ramp slopes as in {\tt fitramp}, we adapt those functions to instead compute the elements of the matrix {\bf A} in Equation \eqref{eq:chisq_element}.  The Legendre polynomial basis is implemented with Numpy's {\tt polynomial} module \citep{numpy1,numpy2}, while linear algebra is also handled by Numpy.  The code is vectorized over pixels and reads to achieve good computational performance.  The software is open-source and freely available.\footnote{\url{https://github.com/RomanSpaceTelescope/SOCReferenceFileCode}}  It can be applied to data from any detector read out up-the-ramp.  

We baseline the computational performance of our implementation by fitting a 10$^{\rm th}$ order polynomial correction to our 285 ramps using a 2023 Macbook Pro.  The total time to process $10^4$ pixels is about 200\,s, scaling to about 100 hours for a full $4096\times4096$ H4RG-10 detector on a single laptop, assuming that we do not compute the condition number of the matrix ${\bf A}$ (the computation of the condition number requires the singular value decomposition, an expensive matrix operation).  Without the condition number, similar amounts of time are spent (1) computing the $g_{i,k}$ for the Legendre polynomials; (2) computing the coefficients of the ${\bf A}$ matrix using Equation \eqref{eq:chisq_element}; and (3) solving Equation \eqref{eq:linear_equation}.  The computation of the $g_{i,k}$ is computationally easier in the standard polynomial basis, reducing the total processing time by $\approx$20-30\%.  We conclude that the approach presented in this paper is a viable method for deriving the nonlinearity correction for the Roman-WFI detectors using large quantities of test campaign data.  

\section{Results on Synthetic Data} \label{sec:syntheticdata}

We test the algorithm described in the previous sections on synthetic data \changes{with Gaussian read noise and Poissonian photon noise}.  We first use 300 ramps of 55 reads each with all ramps using similar count rates and extending to digital saturation.  We assume a sixth-order nonlinearity correction; we use a cubic spline with a large number of knots to numerically invert it.  We then apply this ``inverse nonlinearity correction'' to impose nonlinearity on initially ideal ramps.  We apply read noise and photon noise to the ideal ramps before applying a transformation to impose nonlinearity. \changes{Depending on the source of read noise it might be more appropriately applied to the nonlinear ramps, but the difference vanishes in the limit of small signal where read noise is important.}

For our second test, we create another set of 300 ramps of 55 reads each, but with three different count rates: one extending to 5\% of saturation, one extending to 20\% of saturation, and one extending to saturation.  We will use this range of count rates to uncover biases in our approach and test mitigation strategies.

In the following subsections, we discuss each of these cases separately.  We begin with synthetic data using similar count rates for all ramps.

\subsection{Nearly Equal Count Rates} \label{subsec:equalctrate}

Our first synthetic data test case uses 300 ramps of 55 reads each for 1000 pixels.  We assume a gain of 1.8\,e$^-$/DN and a single-read read noise of 5\,DN; we apply these noise sources to ideal ramps.  We take true count rates from 1450 to 1550\,DN/read; these give final counts of just over 60,000\,DN after applying the inverse of our adopted sixth-order polynomial correction.  We take the pedestal to be 5000\,DN, which then places the ramps into digital saturation at the end.  For each pixel, we fit for the nonlinearity correction using the approach described in Sections \ref{sec:generalapproach} and \ref{sec:formalism}.  

\begin{figure}
    \centering
    \includegraphics[width=0.5\linewidth]{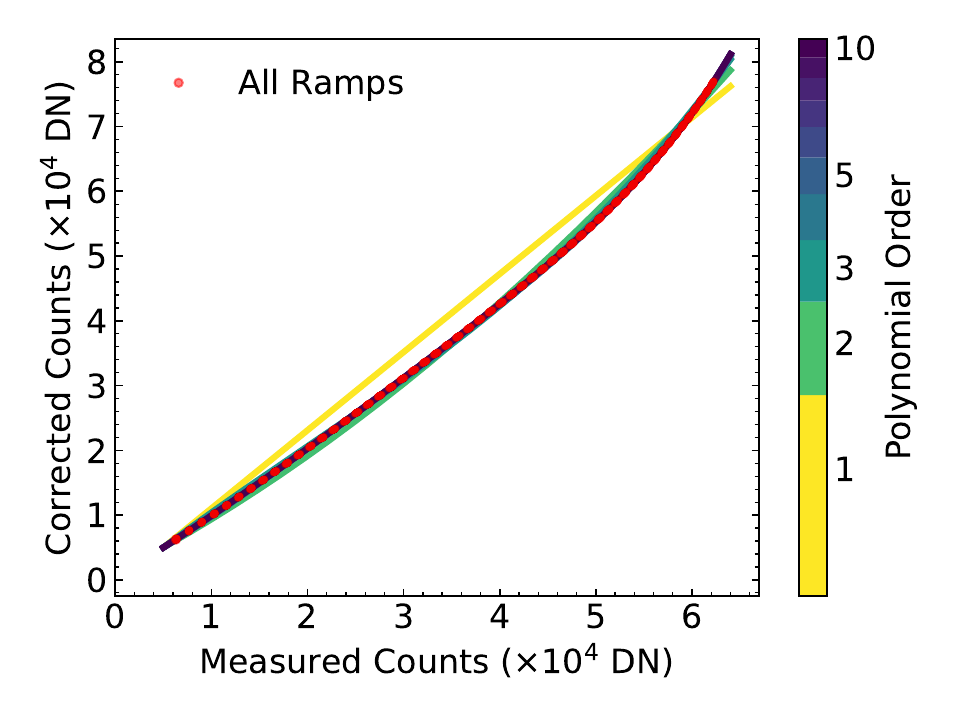}
    \caption{Nonlinearity corrections of orders up to 10 applied to synthetic data that are perfectly corrected (up to read and photon noise) \changes{by} a 6-th order polynomial.  There are 300 ramps all plotted together \changes{with semitransparent red dots for their individual reads}; their count rates vary by $\approx$10\%.  The correction polynomial takes measured counts (shown on the $x$-axis) and outputs corrected counts (on the $y$-axis). }
    \label{fig:fakedata_abs_allsat}
\end{figure}

We begin by showing the measured counts plotted against the corrected counts.  Figure \ref{fig:fakedata_abs_allsat} places the corrected counts on the vertical axis and the measured counts on the horizontal axis.  The best-fit corrections are shown for polynomial orders up to 10.  Our use of measured and corrected counts for the two axes enables us to plot many ramps of varying count rates together.  

\begin{figure}
    \centering
    \includegraphics[width=0.495\linewidth]{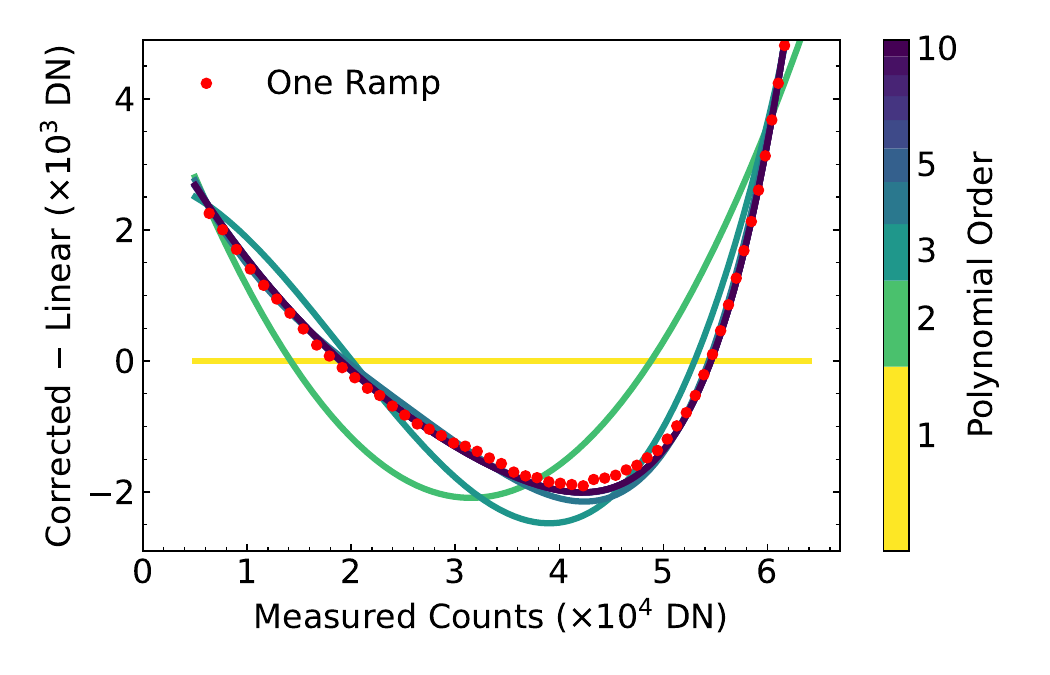}
    \includegraphics[width=0.495\linewidth]{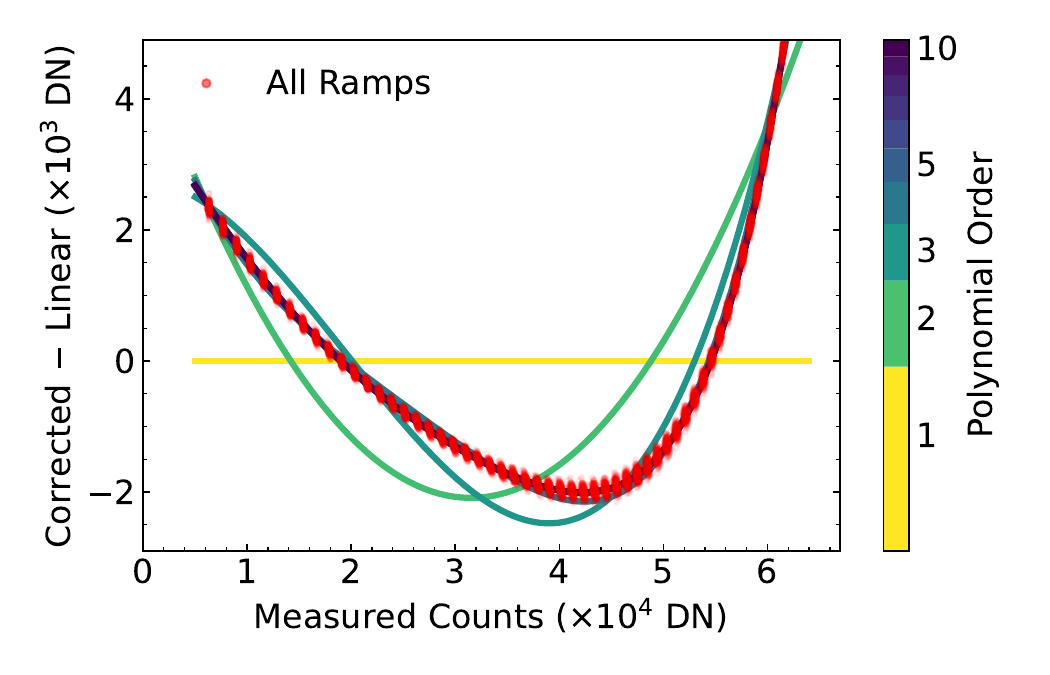}
    \caption{The same data as in Figure \ref{fig:fakedata_abs_allsat}, but plotted relative to a linear fit to better show variations.  The left panel shows one ramp; the right panel shows all 300 ramps used to derive the nonlinearity correction.  The ramps have a 10\% range of count rates, so to that individual reads cease to become visually distinguishable after around read 10.}
    \label{fig:fakedata_rel_allsat}
\end{figure}

Figure \ref{fig:fakedata_rel_allsat} shows the same data as Figure \ref{fig:fakedata_abs_allsat}, but with the corrected counts measured relative to the linear fit (yellow line in Figure \ref{fig:fakedata_abs_allsat}).  The left panel shows a single ramp with the impact of photon noise clearly visible.  This noise causes correlated residuals from even the highest-order corrections.  The right panel of Figure \ref{fig:fakedata_rel_allsat} shows all reads of all ramps plotted together with semitransparent points; the correlated nature of photon noise becomes imperceptible.  The use of a range, albeit a narrow one, of count rates fills in the measured counts on the horizontal axis.  

Figure \ref{fig:fakedata_rel_allsat} shows that the data require a polynomial correction of order $\geq$4.  To determine the degree of the correction, and to verify the statistical robustness of our approach, we measure the average improvement of $\chi^2$ across our 1000 simulated pixels.  The $\chi^2$ value is computed for all ramps for a given pixel.  With 300 ramps, 55 reads per pixel, a free ramp-by-ramp reset value, a free ramp-by-ramp slope, and a handful of saturated reads, we have $\approx$15,000 degrees of freedom per pixel.  We lose one degree of freedom for each additional order in our polynomial fit.  In this case, the data were derived by numerically inverting a sixth-order correction polynomial; there should be no benefit to a correction of order higher than six.  The improvement in $\chi^2$ with each additional order should then be $\approx$1 as the additional free parameter begins to fit nothing but noise.  

\begin{figure}
    \centering
    \includegraphics[width=0.5\linewidth]{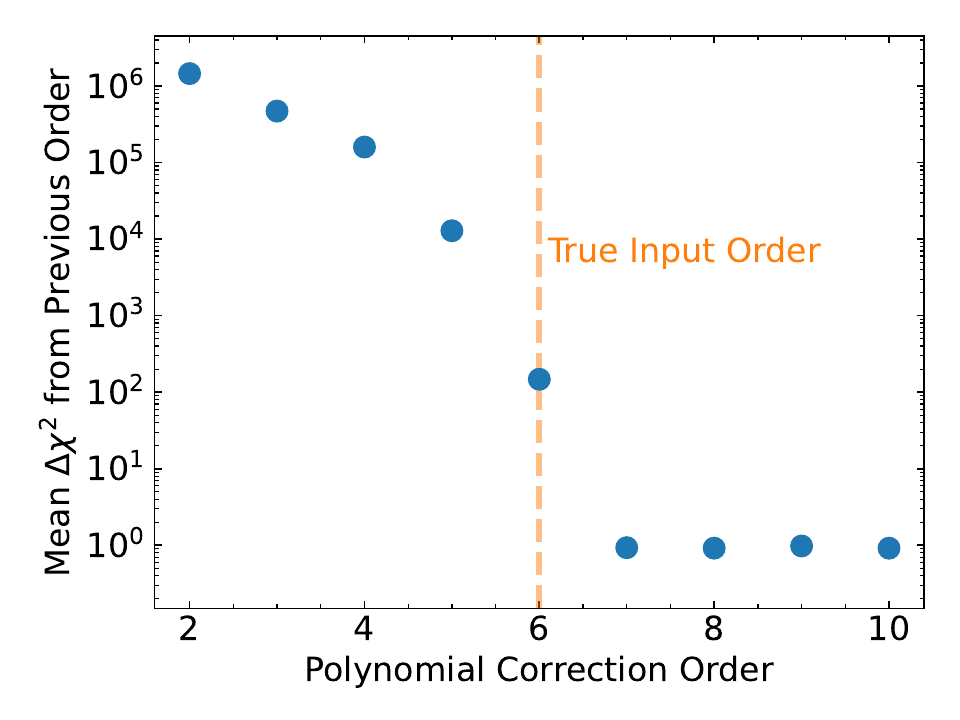}
    \caption{The incremental improvements in $\chi^2$ that can be achieved with increasing polynomial correction order, shown on a logarithmic scale.  The $\chi^2$ value is computed for 300 ramps of 55 reads each.  In this case, the synthetic data are perfectly corrected, up to read and photon noise, by a sixth order polynomial.  Improvements in $\chi^2$ past this point are $\approx$1 per additional order as expected from noise in the absence of signal.}
    \label{fig:chisq_improvements}
\end{figure}

Figure \ref{fig:chisq_improvements} shows the improvements in $\chi^2$ with each additional order of the polynomial fit.  The improvement is very large for the initial increases in the order of the correction.  Visually, this corresponds to the much better visual fits in Figure \ref{fig:fakedata_rel_allsat} with an order 2 compared to an order 1 correction, etc.  The improvements drop significantly with order, before abruptly settling on $\Delta \chi^2 \approx 1$ past order 6.  This corresponds to the ideal behavior of an estimator.  So long as our statistical assumptions of read and photon noise are satisfied, it confirms the effectiveness of our formalism for many ramps of varying count rates that all extend to full well.  

As a final test, we measure the deviation from the input nonlinearity correction for each of the 1000 ramps.  This deviation is measured relative to the linear term, so it is zero by construction at the pedestal, or the mean reset value.  We measure it as a fractional deviation.  An interpretation of this deviation is the fractional error in the ratio of the inferred flux of a faint star (that only reaches a small fraction of full well) to that of a bright star.  Such an error, if it has a nonzero mean value across the 1000 pixels, will mimic count rate dependent nonlinearity \citep{Bohlin+Lindler+Riess_2005}.

\begin{figure}
    \centering
    \includegraphics[width=0.5\linewidth]{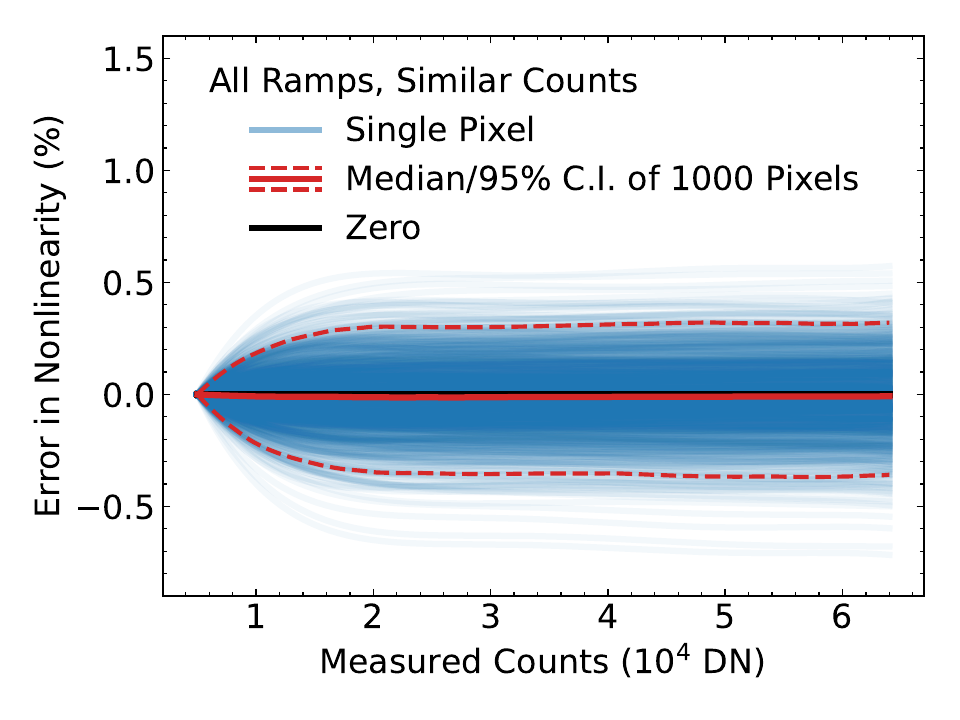}
    \caption{Percent error in the derived nonlinearity correction as a function of counts for synthetic data where the true (sixth-order) correction is known.  Each line represents the fit for 300 ramps of 55 reads each; the dashed lines show the empirical 95\% confidence interval.  With 300 ramps per pixel of similar count rates, the approach is nearly unbiased.}
    \label{fig:bias_fakedata_allsat}
\end{figure}

Figure \ref{fig:bias_fakedata_allsat} shows the results, with each of 1000 pixels plotted as a semitransparent line.  They are all zero by construction at the pedestal value of 5000\,DN and diverge in both directions, with some pixels (by chance) overcorrecting and other pixels undercorrecting the true nonlinearity.  The median correction, shown by the red line, is very close to zero across the entire dynamic range, with residual biases at a level $\ll$0.1\%.  \changes{The shape of the curves reflects a well-determined nonlinearity correction at high count levels (a nearly flat line) with an uncertain anchor to the slope at the pedestal value.  Varying this slope---the linear term in the nonlinearity correction---will vertically offset the residual at high counts.}

\subsection{Widely Varying Count Rates} \label{subsec:varyingcountrates}

We now conduct a series of tests using 300 ramps, with 100 each of 50-60 DN/read; 200-230 DN/read; and 1300-1400 DN/read.  With 55 reads, these reach $\approx$5\%, $\approx$20\%, and 100\% of our adopted saturation limit.  We assume the same sixth-order nonlinearity correction adopted in the previous subsection.  

\begin{figure}
    \centering\includegraphics[width=0.5\textwidth]{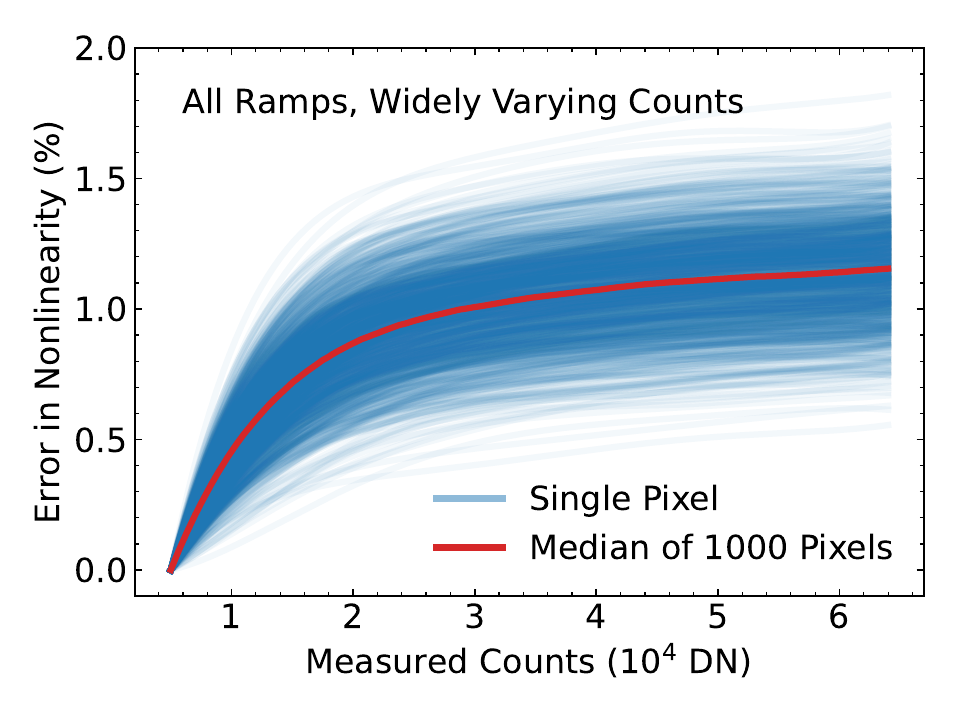}
    \caption{Similar to Figure \ref{fig:bias_fakedata_allsat}, but for the case where the count rates of the ramps vary by a factor of $\approx$20.  There is now a $\approx$1\% bias: the fitting process overestimates the correction at high counts relative to low counts.  The reason for this is discussed in the text. \label{fig:bias_varctrate}}
\end{figure}

As a first test, we repeat the nonlinearity fit described in the previous section using all ramps simultaneously.  Figure \ref{fig:bias_varctrate} shows the results.  In contrast to Figure \ref{fig:bias_fakedata_allsat}, a strong bias is now present, with a $\approx$1\% systematic error in the pixel's nonlinearity correction across its dynamic range.  If present in the nonlinearity correction applied to real data, this will introduce apparent count-rate-dependent nonlinearity.

The bias in Figure \ref{fig:bias_varctrate} may be understood as the \changes{unfortunate interaction of a number of facts.}  First, we are evaluating the likelihood on the corrected, rather than the raw, count rates.  \changes{This is, in fact, the assumption used to generate the data, and will reflect the properties of real data if nonlinearity is due to effects other than decreased photon conversion efficiency.  Fitting the raw counts as a function of time would derive the inverse of the linearity correction, not the linearity correction that we seek. Because the likelihood is evaluated on the corrected counts, a nonlinearity correction with a small value for the linear term will reduce the scatter (and hence $\chi^2$) in ramps with low count rates.  If these ramps have lower uncertainties and a higher density of samples, the correction at low count rates can be underestimated.  The correction at higher count rates will be worse as a consequence, but if the uncertainties are larger and the density of measurements is lower, the $\chi^2$ penalty can be smaller than the gain from an underestimated linear term. }

\changes{To better establish the cause of the bias seen in Figure \ref{fig:bias_varctrate}, we perform a number of tests.  First, we fix the covariance matrices to their correct values for each ramp.  This has a negligible impact on the bias.  Next, we iteratively scale the covariance matrix to an estimate of the raw counts as follows.  Given a nonlinearity correction, we can compute the ratio of the raw and linearized count rates and scale the covariance matrix read difference-by-read difference by the square of this ratio.  This procedure does reduce the bias, but only slightly.  To verify that our use of a Gaussian likelihood to model Poisson noise is not the problem, we regenerate synthetic ramps using the Gaussian distribution for photon noise.  The resulting bias is indistinguishable from that seen in Figure \ref{fig:bias_varctrate}.   Finally, we fix the covariance matrix to be the same for all ramps, though it is now an incorrect statistical description of the data.  We use fixed covariance matrices of either pure read noise or pure photon noise.  }

\begin{figure*}
    \includegraphics[width=\textwidth]{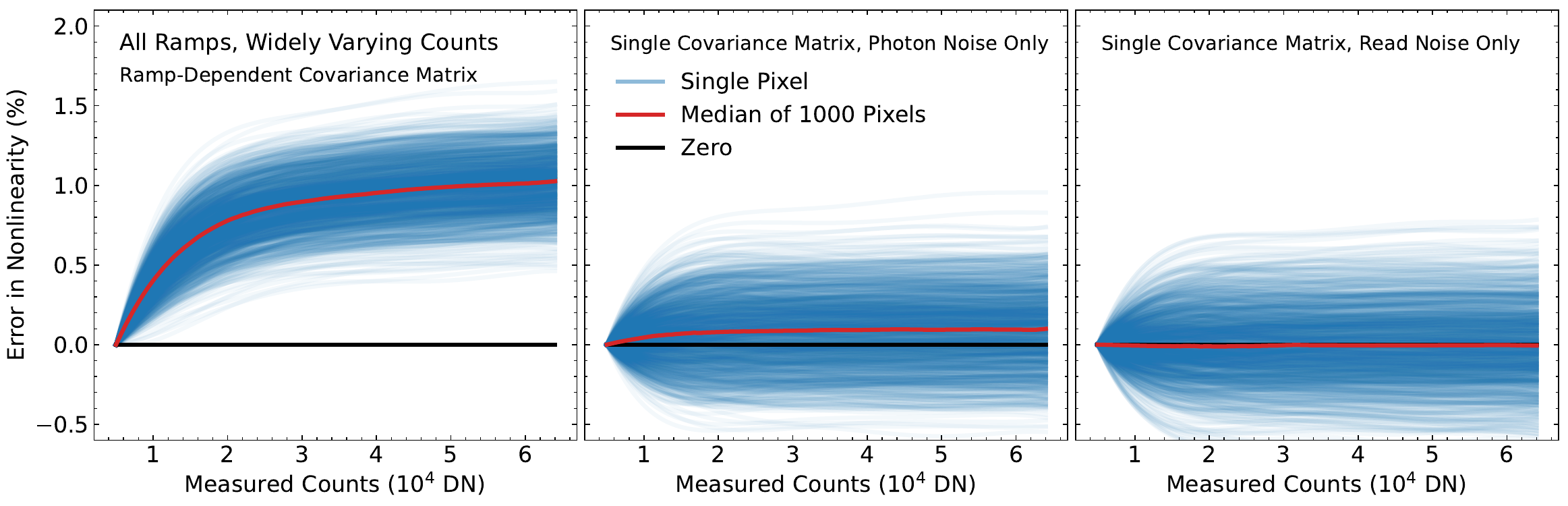}
    \caption{\changes{Change in the bias shown in Figure \ref{fig:bias_varctrate} under three of the tests described in the text.  In the left panel, the covariance matrices are rescaled to simulate fitting to the raw rather than corrected counts.  The bias is decreased relative to that seen in Figure \ref{fig:bias_varctrate}, but only slightly.  In the right two panels, the same covariance matrix is used for all ramps, despite the fact that it is then no longer a correct description of the data.  Using a fixed photon noise-like covariance matrix greatly reduces the bias, while using a read noise-like covariance matrix visually elimininates it.  These results strongly suggest that the bias is partially due to the different uncertainties in ramps of low and high count rates, combined with the correlated nature of the errors.} \label{fig:bias_comparison_all}}
\end{figure*}

\changes{Figure \ref{fig:bias_comparison_all} shows the results of three of the tests described above.  The left panel establishes that scaling the covariance matrix read difference by read difference to mimic fitting the raw data does reduce the bias, but only modestly, by about 10\%-15\%.  Using a fixed covariance matrix for all ramps, however, dramatically reduces the bias.  In the case of a covariance matrix with the structure of read noise (right panel), the bias is all but eliminated. Using a covariance matrix with the structure of photon noise more accurately describes the ramps with high count rates, but this does not fully remove bias in the derived nonlinearity correction.  Neither of these covariance matrices is a correct statistical description of the data.  The results shown here, combined with the lack of significant biases in Figure \ref{fig:bias_fakedata_allsat} suggest that a combination of factors is to blame for the bias: a changing density of measurements at different count levels, correlated noise, and larger uncertainties at higher count levels. }

\begin{figure}
    \centering\includegraphics[width=\textwidth]{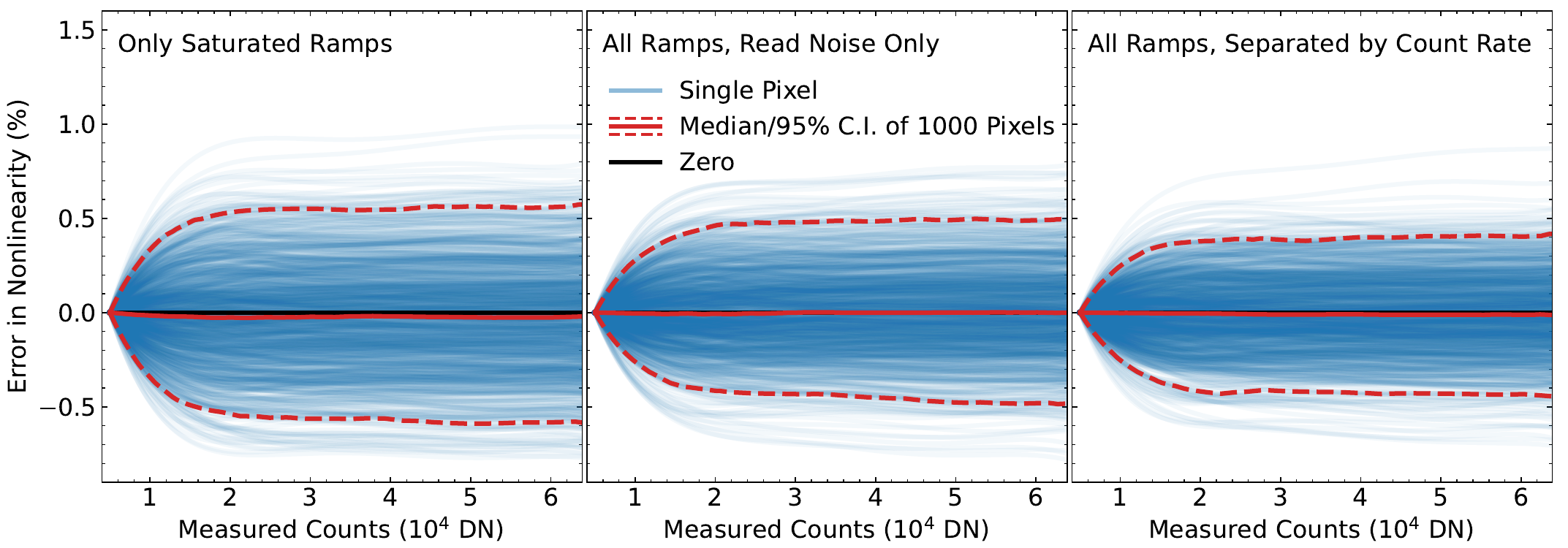}
    \caption{Results of three approaches to mitigating the bias in Figure \ref{fig:bias_varctrate}.  The left panel shows the result using only the 100 ramps (of 300 total) that extend to saturation and that have similar count rates.  The middle panel shows the results if approximating the covariance matrix as read noise only, with no dependence on the inferred count rate.  The right panel shows the results of separately fitting the ramps in groups of similar count rates and then combining them.  All three approaches greatly reduce any bias, with strengths and weaknesses that are discussed in the text.  \label{fig:bias_fakedata_modelramps}}
\end{figure}

\changes{To derive a nonlinearity correction for real data, we must overcome the biases discussed above.  One approach is to restrict the analysis described in Section \ref{sec:formalism} to ramps that reached saturation, about one-third of the ramps in our case.} This result, shown in the left panel of Figure \ref{fig:bias_fakedata_modelramps}, closely corresponds to Figure \ref{fig:bias_fakedata_allsat}, albeit with less data.  The use of only a subset of the available data sacrifices signal-to-noise along with resolution at low count rates, but removes nearly all of the bias seen in Figure \ref{fig:bias_varctrate}.  

\changes{We suggest two other approaches} to overcome the bias in the nonlinearity correction. First, \changes{as in the right panel of Figure \ref{fig:bias_comparison_all}, we may} simply ignore photon noise in the covariance matrix.  This removes one of the contributors to the bias: that disagreements at high signal are weighted less than disagreements at low signal.  Our covariance matrix is less faithful to the actual statistics of the problem, and we can no longer interpret $\chi^2$ as a log likelihood.  This limits our ability to derive confidence intervals on a nonlinearity correction, although it is not an obstacle to deriving the correction itself.  The \changes{residuals from this approach, shown in the right panel of Figure \ref{fig:bias_comparison_all}, are replotted in the middle panel of Figure \ref{fig:bias_fakedata_modelramps}}. The bias is now too small to measure, and the scatter of the result across 1000 synthetic ramps is slightly improved from the case of only using saturated ramps.  Importantly, we can now make use of all of the data, and ensure that our correction is appropriate at both high and low count levels.  

For our \changes{final} approach to remove bias, we address the problem that arises from combining ramps of different count rates.  In the previous section, we demonstrated that the bias in the nonlinearity correction is very small when all ramps reach saturation. In order to add the information supplied by the the ramps extending to only a small fraction of saturation, we must mitigate their tendency to bias the fit.  We proceed by treating ramps of similar count rates as separate groups before merging the results.  For the synthetic data described here, that means that we split our data into three groups of 100 ramps each.  We fit a sixth-degree polynomial for each nonlinearity correction (the true model).  Each group of ramps has a small, but nonzero, range of true count rates.  Fitting each of these groups of 100 ramps results in three nonlinearity corrections for each pixel.  

After deriving three nonlinearity corrections per pixel, each with its own dynamic range, we must combine them.  We construct three model ramps to represent the model correction at each count level.  We use a cubic spline with a large number of knots to numerically invert each of our three best-fit nonlinearity corrections, and use model ramps extending up to the median count rate among the input ramps used.  This model ramp is noiseless in the sense that it is perfectly corrected by the best-fit polynomial nonlinearity correction.  The only noise present comes from photon noise and read noise that resulted in imperfect fitted nonlinearity corrections.

The three such noiseless ramps that we construct will not quite be consistent with a single nonlinearity correction, as each ramp had its own realizations of read noise and photon noise.  We therefore use the algorithm described in Section \ref{sec:formalism} to derive the best-fit nonlinearity correction for these three model ramps.  In each case, we assume apply the algorithm assuming that the noise is still given by read and photon noise, albeit with the read noise scaled down by the square root of the number of contributing ramps $n$, and the gain scaled up by a factor of $n$.  This determines the relative weights applied to each nonlinearity correction.

The right panel of Figure \ref{fig:bias_fakedata_modelramps} shows the results of this approach.  The pixel-to-pixel scatter in the derived nonlinearity corrections is reduced relative to the cases of only using the ramps that extend to saturation (left panel) and the case of neglecting photon noise (middle panel).  The scatter in the nonlinearity corrections is almost to the level seen in the case of 300 reads all extending to saturation (Figure \ref{fig:bias_fakedata_allsat}).  In other words, these ramps reaching lower count rates improve the nonlinearity correction almost as much as if they extended all the way to saturation.  The bias is also dramatically reduced relative to that seen in Figure \ref{fig:bias_varctrate}; the median error among 1000 pixels is $\ll$0.1\% and appears to be $\lesssim$0.01\%.  

The three approaches described in this section might best be used in differing circumstances.  If most or all available data extend to saturation, then little will be lost by only using ramps that saturate.  If ramps are available at a range of count rates, the approach neglecting photon noise is the most straightforward to apply and offers good performance.  Separately correcting groups of similar ramps and then combining the corrections can offer slightly better performance, but it requires care.  In addition, the algorithm of Section \ref{sec:formalism} is more vulnerable to numerical difficulties in this case.  For the data used in this section, combining the correction from groups of ramps begins to fail at a polynomial order of $\approx$10, while the approach using read noise only is well-behaved up to an order of $\approx$20.  

Based on the results of this section, our recommendation is to apply the algorithm of Section \ref{sec:formalism} directly if all available ramps extend to saturation.  If ramps are available with a range of count rates and integration times, we generally recommend using a covariance matrix that ignores photon noise (e.g.~by assuming infinite electronic gain).  This makes full use of the available data, offers \changes{essentially} unbiased corrections, and achieves nearly the same precision of processing the ramps in similar groups.

\section{Comparison to a Legacy Approach} \label{sec:comparison_canipe}

In this section we compare the performance of the algorithm developed in this paper to that presented in \cite{2017jwst.rept.5167C}.  In that case, the true count rate $b$ is first estimated by fitting a low-order polynomial to the first few reads and taking the linear term at the reset value.  The correction is then obtained by minimizing
\begin{equation}
    {\cal C} = \sum_i \left( b \cdot i - \sum_{j=0}^{N} a_j (y_i)^j \right)^2 \label{eq:canipe}
\end{equation}
with respect to the $\{a_j\}$, where $i$ is the read number and $N$ is the order of the fit.  The final correction is given by $\{a_j/b\}$, with the nonlinear part of the correction given by the terms with $j \geq 2$.  This approach is nearly equivalent to the one in this paper, albeit with a diagonal approximation to the covariance matrix and with difficulty in treating multiple ramps simultaneously.  \changes{A diagonal covariance matrix mitigates bias as shown in Section \ref{subsec:varyingcountrates} but it sacrifices signal-to-noise ratio in the case of a single ramp and makes the $\chi^2$ improvement test shown in Figure \ref{fig:chisq_improvements} impossible.}

We provide a comparison between the methods limited to the case where all ramps have nearly identical true count rates.  In this case, we may average the ramps together read-by-read into a combined ramp before deriving the nonlinearity correction.  This enables us to use Equation \eqref{eq:canipe} straightforwardly.  If the ramps have differing count rates, then this averaging will smooth out actual nonlinearity and will give biased results.

\begin{figure}
    \centering\includegraphics[width=0.8\textwidth]{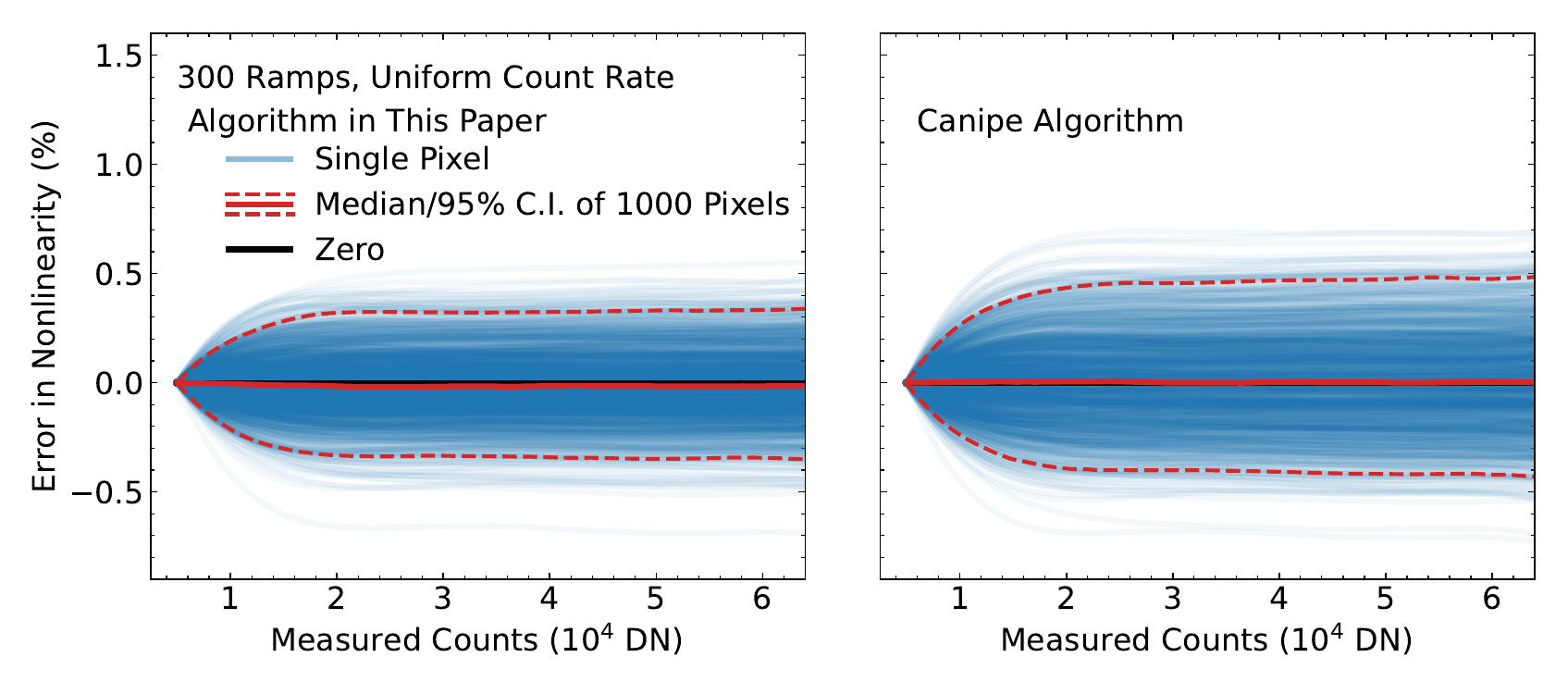}
    \caption{Performance comparison of the nonlinearity correction algorithm in this paper (left) to that of \cite{2017jwst.rept.5167C} (right).  The \cite{2017jwst.rept.5167C} algorithm requires averaging the ramps before deriving the correction, and can only work with a small range of count rates.  With this limitation, both approaches are unbiased or very nearly so, with the algorithm described here offering slightly better precision.    \label{fig:algorithm_compare}}
\end{figure}

Figure \ref{fig:algorithm_compare} shows our results after averaging 300 55-read synthetic ramps with the same read and photon noise properties as in the previous section.  This time, we use the same count rate in each ramp.  The left panel of Figure \ref{fig:algorithm_compare} is nearly identical to Figure \ref{fig:bias_fakedata_allsat}, while the right panel shows the equivalent residuals using the \cite{2017jwst.rept.5167C} algorithm.  This legacy algorithm appears to be free of bias, or very nearly so, when all ramps have the same intrinsic count rate.  The precision of the correction is slightly lower, with the pixel-to-pixel scatter of corrections about 25\% lower with the new algorithm than with the legacy one.  

If the data used for the nonlinearity correction consist of many ramps extending to saturation under (nearly) identical illumination, the \cite{2017jwst.rept.5167C} algorithm performs almost as well as the one presented here.  However, this direct comparison is only possible in the case that we can average ramps together before deriving the correction.  A proper combination of ramps reaching very different fractions of full well, or subject to significantly varying illumination, requires our new approach.  \changes{This situation naturally occurs if we wish to use both flatfield data short of saturation and nonlinearity data extending past saturation, or if we have many exposures taken over long time spans with varying calibration intensities, backgrounds, and/or instrumental throughputs.  }

\section{Results on Roman-WFI Pixels} \label{sec:romanresults}

In this section we apply the methodology developed in the preceding sections on data from the thermal-vacuum (TVAC) testing campaigns of Roman-WFI.  We restrict our analysis to a small region, chosen at random, of a single detector.  We adopt a set of 285 ramps, each composed of 55 nondestructive reads.  About half of these extend to digital saturation at 65535 DN, about one-third are darks, and most of the rest go up to $\approx$10\%-20\% of saturation.  The latter ramps improve the sampling of the detector's response at lower count levels.  In light of the analysis in Section \ref{subsec:varyingcountrates}, we fit for the nonlinearity correction neglecting the contribution of photon noise to the covariance matrix.

\begin{figure}
    \centering\includegraphics[width=0.5\textwidth]{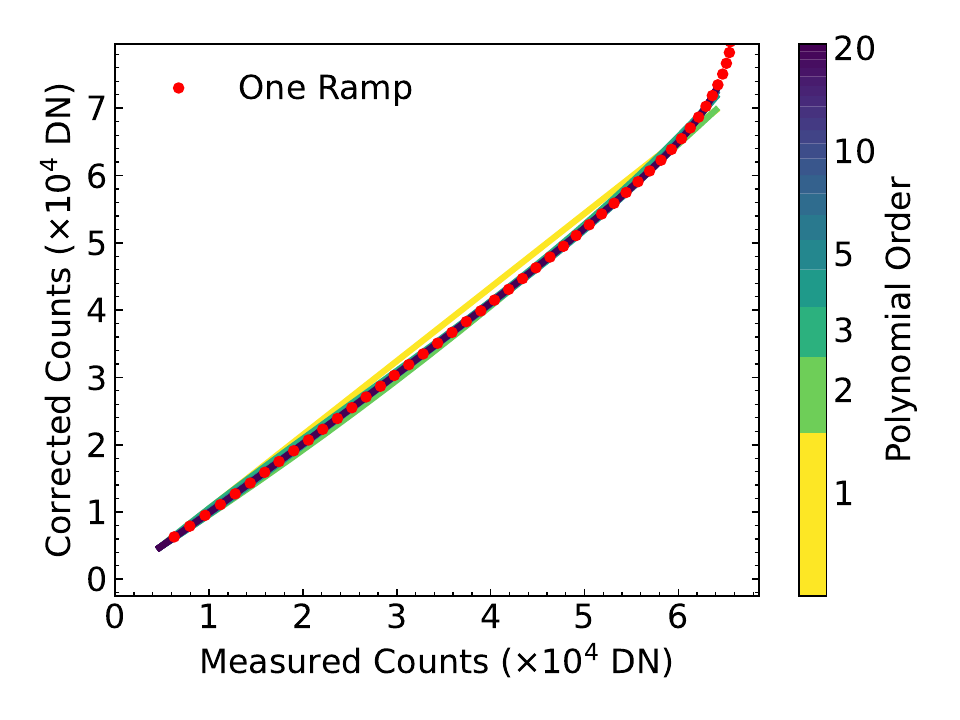}
    \caption{Nonlinearity corrections for a single pixel derived from 186 ramps of 55 reads each, about half of which reach saturation.  The fit covers measured counts up to 64,000 DN, just short of digital saturation, and is done in the Legendre basis neglecting photon noise.  The red points show a single one of these ramps, chosen at random.  The nonlinearity correction is numerically well-behaved up to a polynomial degree of at least 20. \label{fig:correction_demo} }
\end{figure}

Figure \ref{fig:correction_demo} shows a sample correction derived for a single pixel from the 186 illuminated ramps, compared to measurements from one of those ramps shown in red.  The fit is a visually good match to the data, and remains numerically well-behaved up to a polynomial of degree 20 thanks to the use of the Legendre basis set.  The time-like variable is plotted on the vertical axis in Figure \ref{fig:correction_demo}.  For the red points, this time-like value is not measured, but is proportional to time with the proportionality constant $b$ derived in the course of the nonlinearity fit.  The axes are inverted from the usual way of plotting nonlinearity \citep[e.g.][]{2017jwst.rept.5167C} in order to emphasize that we are fitting for the forward correction directly.  Figure \ref{fig:correction_demo} is similar to Figure \ref{fig:fakedata_abs_allsat} albeit for real data and extending to a higher polynomial order.  We note that, since the fit is done in the space of read differences, the offset, or reset value, is never actually fitted.  This is usually treated as a nuisance parameter in ramp fitting because of $kTC$ noise \citep{Brandt_2024a}.  We fit the offset after deriving the nonlinearity correction to provide the best match to the highest-order correction.  This only functions to visualize the correction; it has no impact on the derivation of the nonlinearity correction itself.

\begin{figure}
\includegraphics[width=0.5\textwidth]{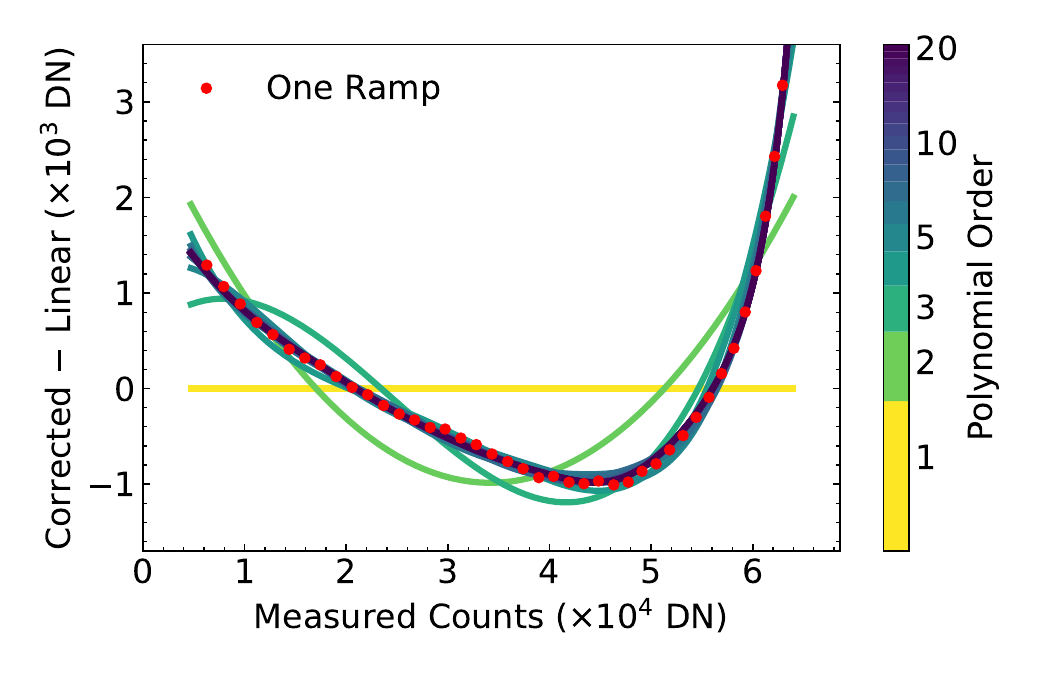} 
\includegraphics[width=0.5\textwidth]{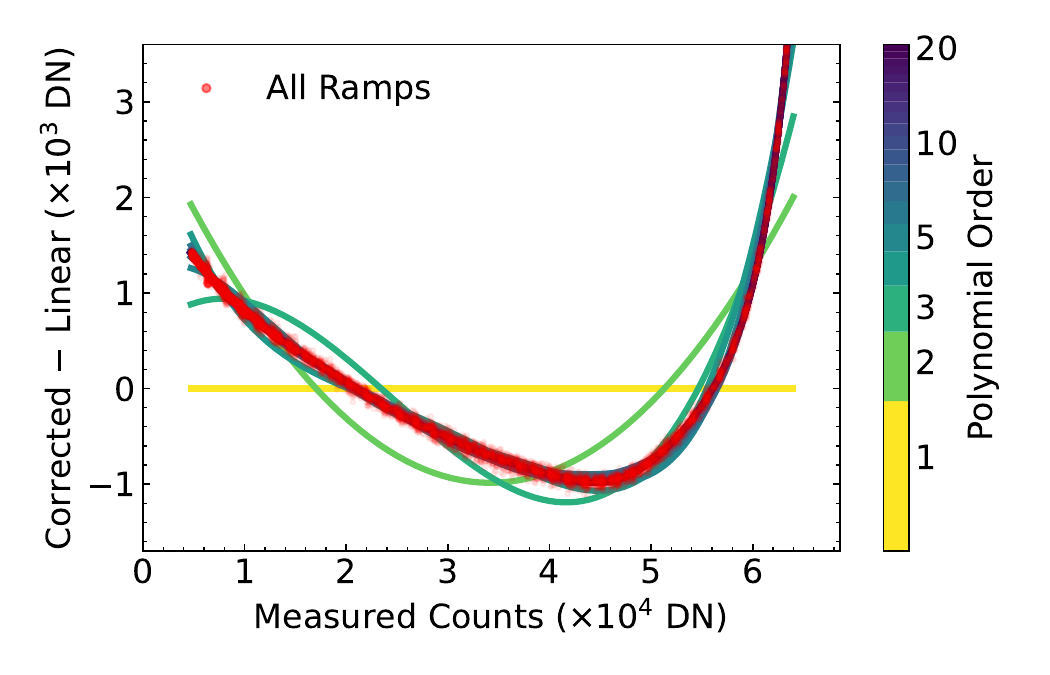}
\caption{Corrected counts relative to a line (i.e.~no nonlinearity correction) as a function of polynomial order.  The left panel shows data from one ramp chosen at random for one random pixel.  The right panel shows all ramps for this pixel.  The correlated residuals that are visible in the left panel arise from different realizations of photon noise that we would like to not be present in a nonlinearity correction.  \label{fig:correction_examples}}
\end{figure}

Figure \ref{fig:correction_examples} shows the same fits as in Figure \ref{fig:correction_demo}, but it plots the residual from applying no nonlinearity correction.  The left panel shows one ramp chosen at random, the same ramp shown in Figure \ref{fig:correction_demo}.  Each ramp has its own realization of photon noise that makes the red points an imperfect match to the high-order curves, even if they are formally good fits.  These correlated residuals are visible because our use of many ramps mitigates the impact of photon noise in any individual ramp.  The right panel of Figure \ref{fig:correction_examples} shows the same residuals, but for all of the ramps simultaneously.  

\begin{figure}
    \centering\includegraphics[width=0.5\textwidth]{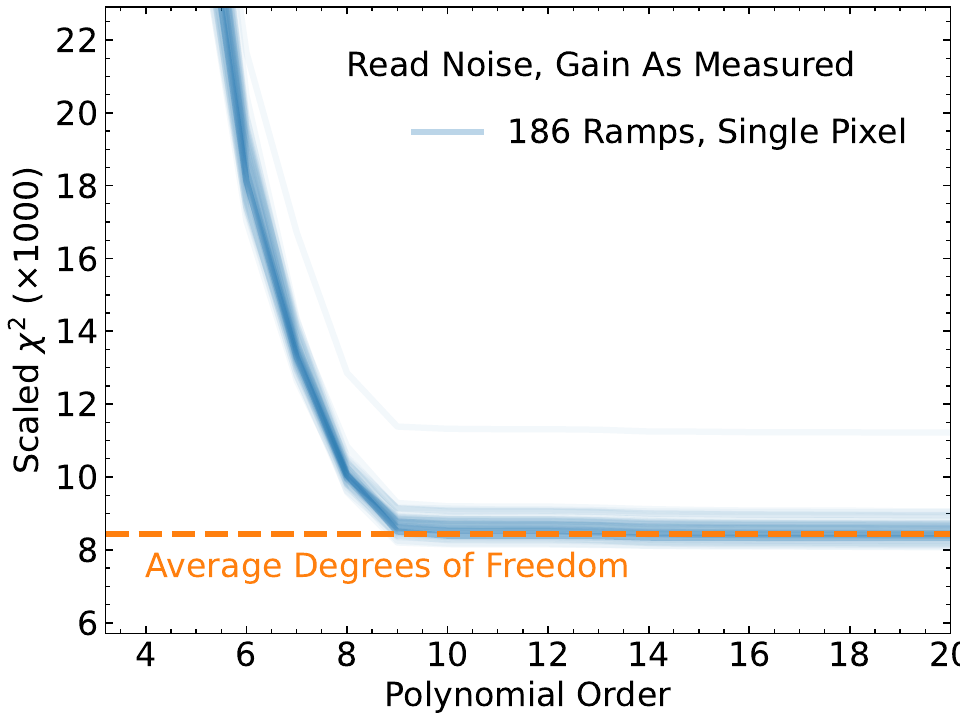}
    \caption{Chi squared values for 186 illuminated ramps of a Roman-WFI detector, each of 55 reads, assuming a gain of 1.8\,e$^-$/DN and a single-read read noise of 5\,DN.  Unlike for Figure \ref{fig:correction_examples}, we do use the full covariance matrix for this exercise.  The dashed orange line indicates the average number of degrees of freedom: the total number of unsaturated reads minus twice the number of exposures (because each one has an unknown reset level and count rate).  Individual lines are for 50 random pixels, all showing a clear flattening of $\chi^2$ at 9$^{\rm th}$ order.  \label{fig:chisq}}
\end{figure}

Figure \ref{fig:chisq} shows $\chi^2$ as a function of polynomial order for 50 random pixels.  Each $\chi^2$ value is the total from 186 illuminated ramps each of 55 reads.  For this exercise, we use the full covariance matrix: our goal is not to derive a bias-free correction but to explore how the likelihood improves with polynomial order.  We fit for a free slope for all but one of these ramps, the fitting method only uses read differences rather than the reads themselves, and some of the reads are saturated and therefore not used.  As a result, a formally good $\chi^2$ value is $\approx$8500.  Our use of representative values for the gain and read noise gives formal $\chi^2$ values that closely correspond to expectations for most pixels. 

For all 50 pixels, Figure \ref{fig:chisq} shows a prominent flattening in $\chi^2$ at order 9.  Below this level, $\chi^2$ increases sharply (and off the plotted range) with decreasing correction order.  For orders higher than 9, increasing order gives little improvement in the quality of the correction.  This improvement is still larger than expected for fitting pure noise, and could potentially be due to a combination of the effects discussed in Section \ref{subsec:varyingcountrates}, high-order nonlinearities, and/or deviations from the assumptions of the classic nonlinearity correction are present.  Some of this could be due to a phenomenon called {\it integral nonlinearity}, in which the analog-to-digital converter induces strongly varying (albeit low amplitude) nonlinearities across the entire signal range \citep{Loose+Smith+Alkire+etal_2018}.

In practice, Figure \ref{fig:chisq} can be computed for a random set of pixels across the Roman-WFI detectors.  One can then look for a natural break point, an ``ankle'' at which $\chi^2$ flattens, to determine the degree of polynomial to be used for the nonlinearity correction.  

\section{Conclusions} \label{sec:conclusions}

In this paper we have derived a \changes{new algorithm to compute} a classic nonlinearity correction.  The approach, building on the methodology of \cite{Brandt_2024a}, uses the tridiagonality of the covariance matrix of the read differences to make the algorithm computationally efficient.  We adopt a polynomial basis for the nonlinearity correction and solve for it directly, rather than taking an intermediate step of first approximating the initial count rate or of fitting for the measured counts as a function of time (an inverse nonlinearity correction).  

The algorithm presented here has a computational cost that is linear in the total number of reads used and weakly nonlinear in the order of the correction polynomial.  Numerically, it is well-behaved even up to a polynomial correction of order 20.  The computational cost can be significant for large amounts of data, but it remains tractable for many hundreds of ramps ($\gtrsim$10$^4$ individual reads) on the $4096 \times 4096$ pixels of a Roman-WFI H4RG-10 detector, taking $\sim$100 hours on a single 2023 Macbook Pro laptop with $\approx$15,000 total reads per pixel.

We show that our proposed approach can be significantly biased if the ramps cover a wide range of count rates.  For most use cases, we suggest deriving the nonlinearity correction with a covariance matrix neglecting photon noise.  In synthetic data sets, this reduces the bias to unmeasurable levels.  The likelihood that we derive via $\chi^2$ may still be used to compute the goodness-of-fit of the correction.  

Finally, we demonstrate the new algorithm on a random set of pixels from one detector of Roman-WFI.  We find good formal agreement between the corrected counts and the model for a polynomial order $\geq$9, with a sharp degradation in this agreement for lower orders.  We conclude that our approach is suitable for determining the necessary correction order, and for deriving the nonlinearity correction of large format detectors from heterogeneous data sets.  

{\it Software}: scipy \citep{2020SciPy-NMeth},
          numpy \citep{numpy1, numpy2},
          matplotlib \citep{matplotlib},
          Jupyter (\url{https://jupyter.org/}).

\begin{acknowledgements}
  I thank the referee, Robert Lupton, for suggestions that significantly improved the manuscript.
\end{acknowledgements}

\end{document}